\documentclass{an}
\usepackage{graphicx}
\usepackage{times}
\Volume{01}              
\Year{2002}              
\Month{10}               
\Pagespan{223}{227}      

\begin{document}
\lhead[\thepage]{L. \v C. Popovi\'c et al.: Observation of H II region in
LEDA 212995}
 \rhead[Astron. Nachr./AN~{\bf XXX} (2004) X]{\thepage}
\headnote{Astron. Nachr./AN {\bf 32X} (2004) X, XXX--XXX}

\title{H II  emission line region in LEDA 212995, a small neighboring
galaxy of Mrk 1040}

\author{L.\v C. Popovi\'c\inst{1,2}, E. G.
Mediavilla\inst{3}, E. Bon\inst{1},
 D. Ili\'c\inst{4} \and G. Richter\inst{2}}

\institute{
Astronomical Observatory, Volgina 7, 11160 Belgrade
74, Serbia
\and
Astrophysikalisches Institut Potsdam, An der Sternwarte 16,
D-14482 Potsdam, Germany
\and
Instituto de Astrof\'\i sica de Canarias
     E-382005, La Laguna, Tenerife, Spain
\and
Department of Astronomy, Faculty of Mathematics, University of
Belgrade,  Studentski trg 16, 11000  Belgrade, Serbia} 

\date{Received ; accepted}

\abstract{
 We present here new spectroscopic observations  of  Mrk 1040 and
LEDA 212995 (Mrk 1040 companion) obtained with the Isaac
 Newton Telescope (INT). The intensity ratios and widths for the narrow
emission lines found in LEDA 212995 are typical of H II regions.  The 
 red-shift (0.0169$\pm$0.00015) of the object derived from these emission
lines is  very close to the red-shift of Mrk 1040
 (z=0.01665). The weak narrow and broad  absorption lines were detected in
the H$\alpha$ wavelength band of
LEDA 212995 spectra. These absorptions indicate that the companion might
be at
least partly obscured by Mrk 1040. Using this and previous observations we
discuss the possible physical relationship between these two galaxies.
    }

\correspondence{lpopovic@aip.de}

\maketitle   

\section{Introduction}

LEDA 212995\footnote{In NED database this object is noted also  as
UGC 01935 NOTES01 and RX J0228.2+3118, in literature  is often noted as
'Mrk 1040 companion'} is a small galaxy
with dimensions
$0.20' \times
0.10'$ and magnitude $>$19  visually located near the Sy 1 galaxy Mrk 1040
(z=0.01665, Huchra et al. 1999). 
The galactic extinction for this
object was given by
Burstein \& Heiles (1982) and Schlegel
et al. (1998). The other
conversion factors for this galaxy were given by Cardelli et
al. (1989).
\begin{figure*}
\includegraphics[width=16cm]{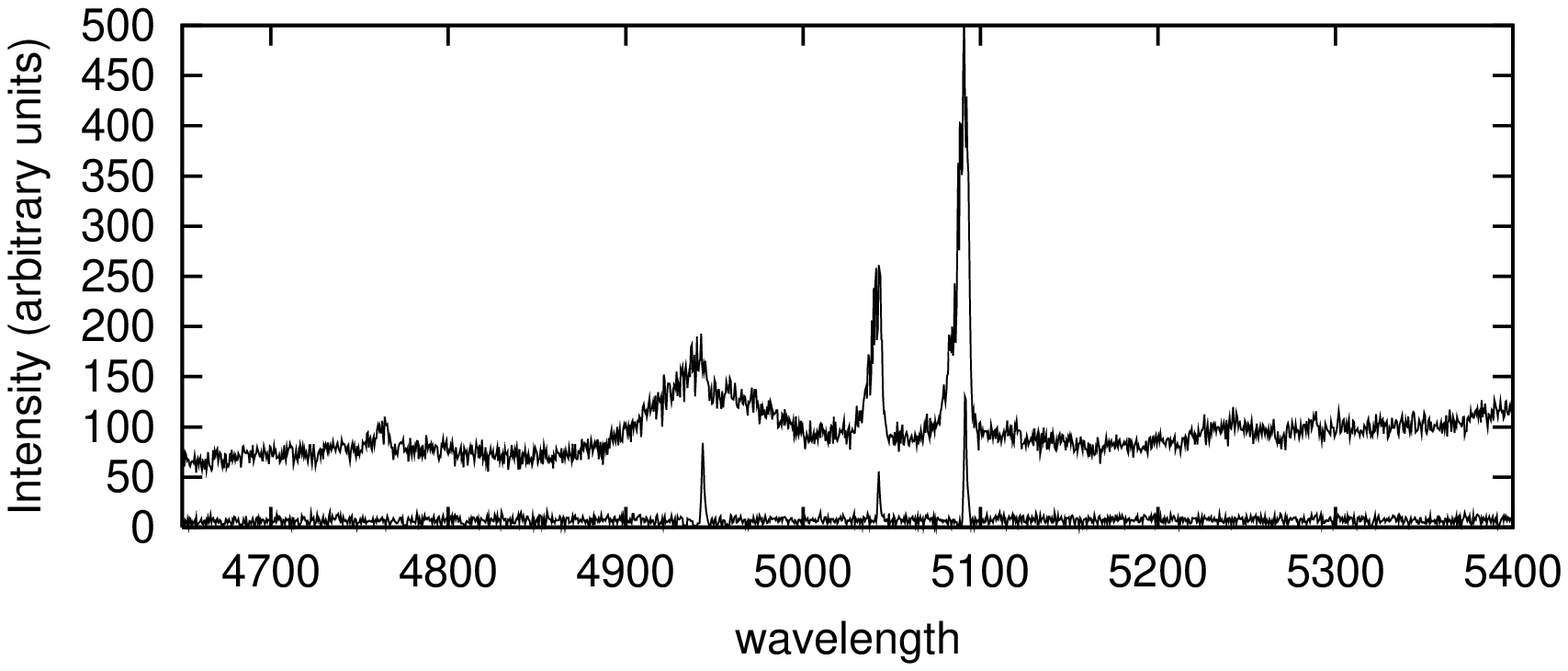}
\includegraphics[width=16cm]{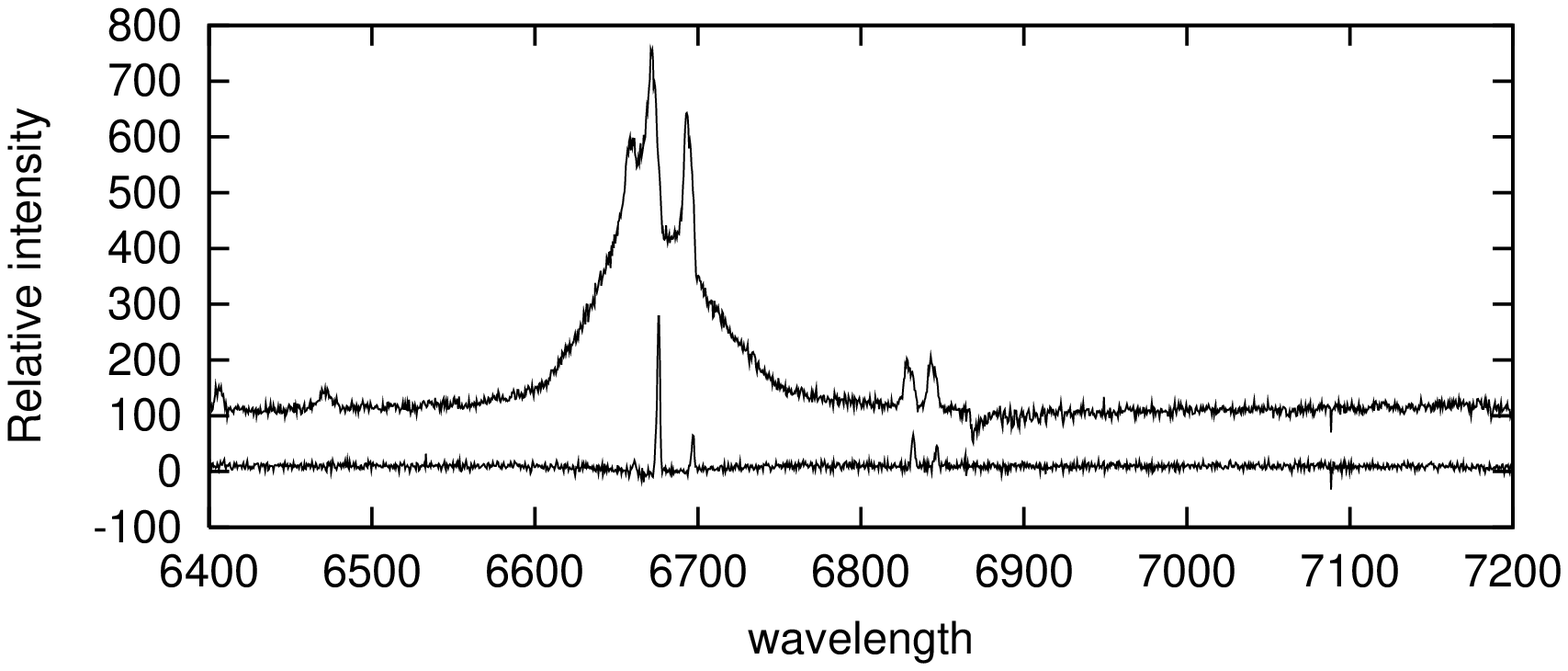}
\caption{ The comparison of the Mrk 1040 and LEDA 212995 spectra in
the H$\beta$  (top) and in the H$\alpha$ (bottom) wavelength range. }
\end{figure*}

\begin{figure}
\includegraphics[width=8.5cm]{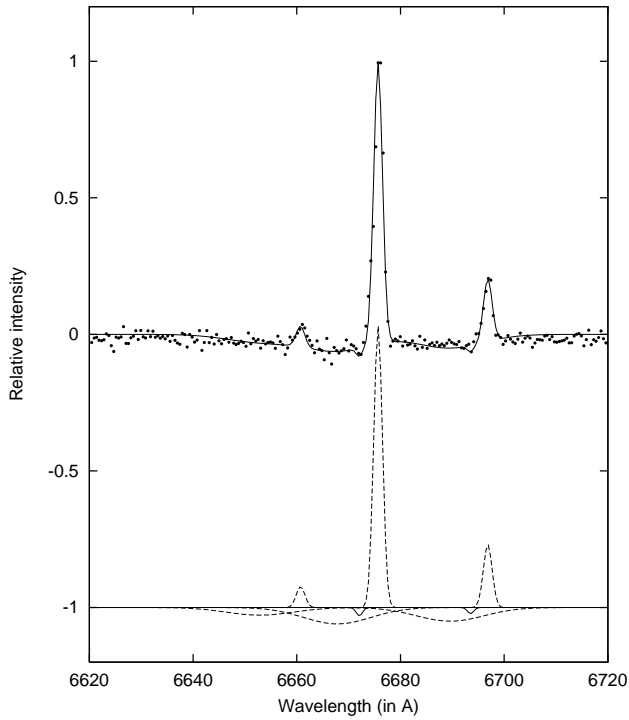}
\caption{Decomposition of H$\alpha$  line region
of LEDA 212995. The 'absorption deep' can be fitted with three broad
absorption (at the bottom) which correspond to the H$\alpha$+[NII] lines
redshifted $z\approx0.0160$. The narrow absorption (solid line) are
 H$\alpha$ and [NII]$\lambda$6583.6 \AA\ with the red-shift of Mrk 1040
($z\approx0.01665$). }
\end{figure}

\begin{figure}
\includegraphics[width=8.5cm]{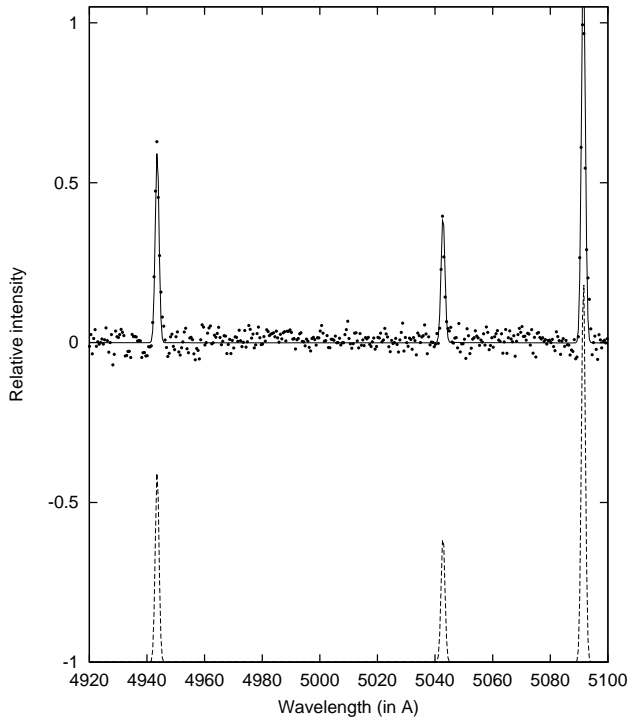}
\caption{Decomposition of  the H$\beta$ line region
of LEDA 212995.}
\end{figure}

\begin{figure}
\includegraphics[width=8.5cm]{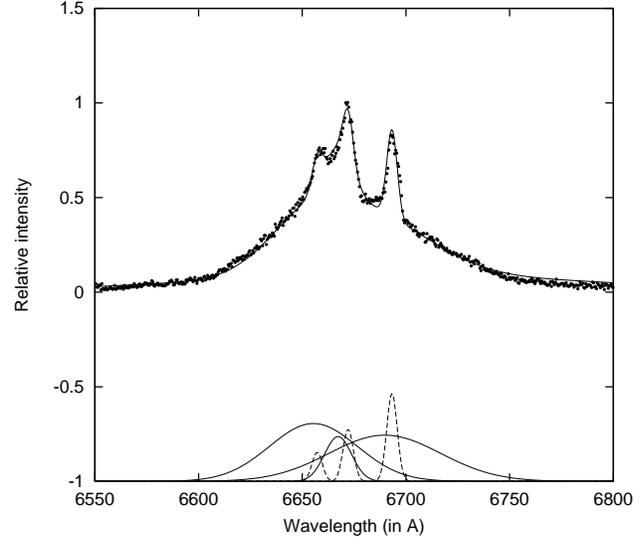}
\caption{Decomposition of Mrk 1040 H$\alpha$ line. The dots
represent
observation, and the solid line is the best fit. The Gaussian components
are shown at the bottom. The dashed lines at the bottom represent the
narrow  lines.} \end{figure}

\begin{figure}
\includegraphics[width=8.5cm]{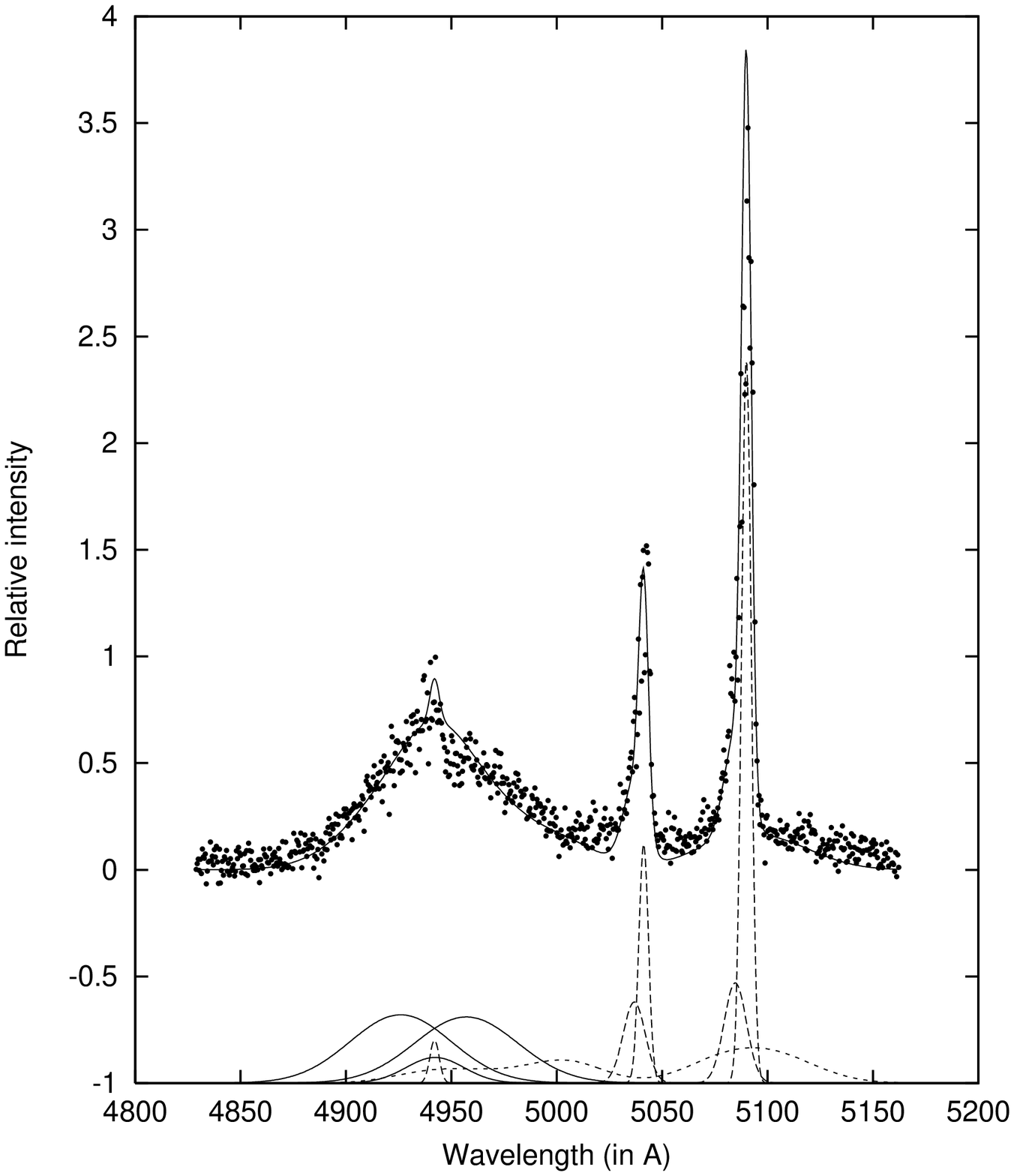}
\caption{The same as in Fig. 4, but for H$\beta$ line.
 The broad dashed line at the bottom represents Fe II template.} \end{figure}

Previous spectroscopical observations  of this object were
reported by Ward \& Wilson (1978), Dahari (1985), Veilleux \&
Osterbrock (1987), Amram et al. (1992) and Keel (1996).
Ward \& Wilson (1987) found that the galaxy emits narrow and 'sharp'
emission lines with red-shift 5070$\pm$60 km/s (0.01691$\pm$0.0002) which
is close to the Mrk 1040 red-shift (4910$\pm$60 km/s,
i.e. 0.01638$\pm$0.0002  according to the
authors). Dahari
(1985) in Table 1, summarized the red-shifts of companions of Seyfert
galaxies where the red-shift of 0.0163 for this object was
given.
 Also, the
spectroscopical observation of this object with CCD
transmission-grism spectrograf at the Cassegrain focus of the Shane 3m   
telescope was analyzed  by Veilleux \& Osterbrock (1987).
Using the emission line ratio they
classified the galaxy as Narrow Emission Line Galaxy (NLRG). Veilleux \&
Osterbrock (1987) found that
systemic red-shift of the object is 0.0160  with the uncertainties
from 0.0002 to 0.0005.
 Assuming this value of the red-shift, the companion should be
closer
to
the observer. Moreover, Afanas'ev \& Fridman (1993) pointed out that an
analysis of the (B-R) color distribution in the galactic disk and the
presence of a distinct dust lane in the disk show that 'the northeast side
of the galaxy is farther away, and the companion, which is bluer than the
disk of Mrk1040, is closer to the observer'. The spectroscopical
observations of  Mrk1040 and LEDA
212995 were given in Amram et al. (1992), where the asymmetry in velocity
field of companion is found and assumed that this asymmetry is due to
interaction of these two galaxies. They found that red-shift of LEDA
212995 is
5100$\pm$ 10 km/s that is in agreement with measurements by Ward \& Wilson
(1978). Keel (1996)
 presented imaging and optical
spectroscopy of a sample paired
Seyfert galaxies and found that LEDA
212995 is about 18 arc seconds to the
north along the Mrk 1040  minor axis.  In this
paper  the  spectroscopic characteristics (red-shift, emission
lines) of the object were not given. 

According to the red-shifts estimates of different authors,  it is
possible  
 to conclude that  galaxies, Mrk 1040 and its companion, are
physically
 related which can be of interest in what respect to the origin of the
 nuclear activity in  Mrk 1040 (see e.g. Corbin 2000).

 We present here new observations of Mrk 1040 and its companion obtained
 with the Isaac Newton Telescope (INT) in the H$\beta$ and H$\alpha$
 spectral regions with the aim of characterize the line emission in Mrk
 1040 companion and discuss the physical relationship between these 
 galaxies.

\section{Observation and data reductions}

The observations were performed on January 24 and 25, 2002 with the 2.5 m
 INT at La Palma. Three exposures of 1400 s in the spectral region of
 H$\alpha$  were obtained on January 24, 2002. Another three exposures of
 1400 s in the H$\beta$ line region were taken on January 25, 2002. The
 Intermediate Dispersion Spectrograph (IDS) and the 235 camera (with chip
 EEV10) in combination with the R1200Y (for the H$\alpha$ wavelength
 region) and R1200B (for the H$\beta$ wavelength region) gratings were
 used. The slit was oriented to obtain simultaneously the spectra  of
 both, Mrk 1040 and its companion. The seeing was $1''.1$ and we used a
 slit $1"$ wide. We estimate a distance between the two objects of about
 18" in agreement with previous measurements (Keel 1996).

CuNe and CuAr lamps were used for wavelength calibration.  Standard
 reduction  procedures including flat-fielding, wavelength calibration,
 spectral response,  and sky subtraction were performed with the help of 
 the
 IRAF software package.   The first step to analyze the emission lines was
 to define and subtract the continuum with DIPSO.

\section{Results and discussion}

Considering that both of the objects were in the same slit we were able to
 see that the red-shift of the LEDA 212995 is slightly higher than
Mrk 1040 one  (see Fig. 1).
In the spectra of this object we easily identify the narrow
H$\beta$, H$\alpha$, [O III]4959,5007 \AA\,; [N II]6548,6583 \AA\,  and
[Si II]6716,6731 \AA\ emision lines.  A very weak and relatively broad
absorption
(like absorption deep) is also
present around 6671 \AA\ (see Figs. 1 and 2).

In order to classify the galactic nucleus, we measured the ratio of the
fluxes of [NII]6584/H$\alpha$ and [OIII]5007/H$\beta$ lines. We found
that $\log\{$[NII]6584/H$\alpha\}$=-0.581$\pm$0.002 and
$\log\{$[OIII]5007/H$\beta\}$=0.169$\pm$0.004. The ratio
of the lines is typical of  HII regions (see e.g. de Burego et
al. 2000). Our measured ratios  are different from
ones given in Veilleux \& Osterbrock (1987), they found that
$\log\{$[NII]6584/H$\alpha\}$=-0.67(-0.68) and
$\log\{$[OIII]5007/H$\beta\}$=0.33(0.36). 

To find the velocity centroids and broadenings as well as to describe the
'absorption deep' in H$\alpha$ wavelength band, we perform a Gaussian fit
to
the emission lines. We used a $\chi^2$ minimalization routine to obtain
the
best fit parameters (see Figs. 2 and 3). To limit the number of free
parameters in
the fit we have set some {\it a priori}  constraints (see e.g. Popovi\'c et
al. 2002, Popovi\'c 2003), taking that  all lines have the same red-shift
and full widths
proportional to their wavelengths. Also, we have linked the intensity ratio
of two [OIII] and [NII] lines according to the atomic values (see e.g.
Wiese et al. 1966); 1:3.03 and 1:2.96, respectively. After subtracting the
continuum, the 'absorption deep' can be clearly seen   in the 
H$\alpha$ wavelength region as well as very weak absorption components
near 
the blue wings of H$\alpha$ and [NII]$\lambda$6583.6 \AA\ 
lines (see Fig. 2). Consequently,
we assumed five  absorption components
more in the fitting procedure for H$\alpha$+[NII] wavelength region.
In H$\beta$ wavelength region, such absorption could not be resolved
due to high level of noise in continuum  (see Fig. 3).

From this analysis we
found that the red-shift  of LEDA 212995 corrected to heliocentric is
$z=0.0169\pm 0.0015$, what is in good agreement with Ward \& Wilson
(1978) and Amram et al. (1992).  We
have found  that the dispersion velocity obtained from emission lines is
about 50 km/s, what is also typical for H II regions.

The absorption components are weak and the parameters from
Gaussian fit are with higher errors than for emission components. From
the fit (see Fig. 2) we found that three broad absorption components may
be present which correspond to H$\alpha$ and
[NII]$\lambda\lambda$6548.1,6583.6 \AA\
lines. The width of the components is $\approx$ 400 km/s and
the red-shift is  $\approx$ 0.016. 
It is interesting that their red-shift and width corresponds to ones of
broader (bluer) [OIII] components ($z\approx 0.0159$,
$w\approx$450km/s) of Mrk 1040 nuclei (see next paragraph).
Additional two weak narrower components
can be detected
only in the H$\alpha$ and [NII]$\lambda$6583.6 \AA\ lines with the
red-shift of 0.01665  and width $\approx$ 30 km/s. 

To check the
obtained red-shift for Mrk 1040 companion, we estimated the red-shift
difference between narrow
spectral lines of both objects. The spectra of Mrk 1040 is more complex
and  to find the velocity  difference,
the $\chi^2$ minimalization routine was applied also to complex Mrk 1040
H$\alpha$ and H$\beta$ lines (see Figs. 4 and 5). We found that systemic
velocities of narrow lines are consistent with cosmological red-shift of
Mrk 1040. Here we should mentioned that narrow
lines of Mrk 1040 are broader and more complex and that uncertainties in
velocity centroid position is higher than in the case of LEDA
212995. Moreover, as one can see in Fig. 4, the [OIII]$\lambda\lambda$
4959,5007 \AA\ lines can be satisfy fitted only if we assume two Gaussian
per the narrow line.
 The velocity difference between
narrow lines of these two galaxies is $z_{\rm LEDA 212995}-z_{\rm Mrk
1040}\approx 0.0004$, that is in agreement  with results
from our previous analysis. 

Next question is: is the LEDA 212995 foreground or background galaxy of
Mrk 1040? The higher red-shift and  'absorption deep' present in the
H$\alpha$ wavelength region indicate that LEDA 212995 may be a background
galaxy. 


 Additional information on this
pair of galaxies can be derived from HST
observations with WFPC2 camera  (for 
more details see Malkan et al. 1998) and from previous observation. The
LEDA 212995 can be seen at the
north edge of the image. By inspection of the
image, one can see that LEDA
212995 shows several distinct clumps and an elongated
structure oriented nearly to the east-west.
An extensive emission in the companion is
observed by Ward \& Wilson (1978) in the low dispersion slit oriented to
the east-west, that  may be present in star-forming regions with such
strong emission lines. From the HST
image, one can clearly see
that LEDA 212995 is projected on spiral arm of Mrk 1040 (northern
side).  Afanas'ev \& Fridman (1993) noted that
 north-east part (where the companion is
projected or seen thought) is receding, and LEDA 212995 has smaller
red-shift with respect
to this part of  galactic disk. Using the Table 2 given in Keel (1996),
where radial
velocities along the major axis  are given, one can find that the velocity
difference in rotation curve for this position in the disk is
(5171-4741)/2=215
km/s. That is higher than relative red-shift difference between two
objects ($\sim$ 120 km/s).
If the companion is background galaxy  one can expect also that
absorption from the stellar disk may be present in the spectra.
But problem is that the absorption components (from
Gaussian analysis) have smaller red-shift than it is expected if absorption
is caused only by north-east (receding) part  of stellar disk.  In any
case 
the north-east part stellar disk may contribute to the observed 
'absorption deep'. 

On the other side,
one can speculate that such small red-shift difference can be caused by
proper motion of LEDA 212995 and that it is only partly obscured by
Mrk 1040 halo (and might be by an extensive NLR), i.e. that the location
of the object corresponds to
the projected distance on the sky between nucleus of   18'' ($\sim 
10Kpc$).  Taking into
account the circular orbit of the companion and mass of Mrk 1040 m$\approx
3\cdot 10^{11}M_\odot$ estimated by Amram et al. (1992), we can conclude
that
projected  velocity of the companion can cause this red-shift
difference. In this case the observed 'absorption deep' might be
intrinsic, i.e. due to absorption of the stars in companion,
reproducing dispersion velocity of stars in LEDA 212995.
Consequently, the companion might be located very close to the nuclei of
Mrk 1040, i.e. practically the companion might be  a part of the Mrk 1040
galaxy.

\section{Conclusion}    

We present our analysis of LEDA 212995 spectra in H$\beta$ and H$\alpha$
wavelength regions. The spectra of this object and Mrk 1040 were observed
simultaneously (both object were in the slit) with INT. From this analysis
we found that: 

i) the intensity ratio of narrow emission lines and their widths
($\approx$ 50
km/s) in the spectra  of LEDA 212995 are  typical for H II regions;

ii) the red-shift inferred from the narrow emission  lines is $z=0.0169\pm
0.0015$ that is in agreement with Ward \& Wilson (1978) and Amram
et al. (1992).  This red-shift imply that LEDA 212995, as was earlier
commented, is really a neighboring galaxy to the Sy 1 galaxy Mrk 1040.

iii) The 'absorption deep' seen
in H$\alpha$ wavelength region indicates that LEDA 212995 is at least
partly obscured by  Mrk 1040, it means that  Mrk 1040 may be a foreground
galaxy of LEDA 212995, or in another case LEDA 212995 might be very close
to Mrk 1040 nuclei. To ascertain this question, further spectroscopic
observations  of the Mrk 1040 companion and north-east part
of the Mrk 1040 stellar disk are needed.

\begin{acknowledgements}
This work is a part of the project P1196  ``Astrophysical Spectroscopy of
Extragalactic Objects'' supported by the
Ministry of Science, Technologies and Development
of Serbia and project                                
P6/88 ``Relativistic and Theoretical
Astrophysics'' supported by the IAC. L\v CP is supported by the Alexander
von Humboldt Foundation through the program for
foreign scholars.
We would like to thank  the anonymous referee for 
very useful comments.
\end{acknowledgements}

{}

\end{document}